\documentstyle[aps,prl]{revtex}
\input  epsf
\begin{document}
\draft

\twocolumn[\hsize\textwidth\columnwidth\hsize\csname@twocolumnfalse%
\endcsname
\title{Simulations of Two-Dimensional Melting on the Surface of a Sphere}

\author{ Antonio P\'erez--Garrido$^a$ and
M. A. Moore$^b$}

\address{$^{a\,}$ Departamento de F\'{\i}sica, Universidad de
Murcia,
Murcia 30.071, Spain}

\address{$^{b\,}$ Theory Group,
    Department of Physics and Astronomy,\\
   The University of Manchester, Manchester  M13 9PL, UK}

\maketitle

\begin{abstract}
{We have simulated a system of  classical particles confined
on the surface of a sphere interacting with a repulsive  
  $r^{-12}$ potential.
The same system simulated on a plane with periodic boundary conditions 
has van der Waals loops in pressure-density plots 
which are usually interpreted as evidence for a first order melting transition,
 but on the sphere such loops are absent.  
 We also investigated the structure factor and from the width of the first
 peak as a function of density we can show that the growth of the correlation 
 length is consistent with KTHNY theory. This suggests  that simulations 
of two dimensional melting phenomena are best performed on the surface of a 
sphere.
}
\end{abstract}

]

Halperin, Nelson \cite{HN78} and Young \cite{Y79}
established several years ago a theory of defect--mediated
melting of two dimensional (2D) crystals. It is based on ideas of Kosterlitz and
Thouless
\cite{KT73}. It is envisaged that the  transition from crystal to
liquid takes place via  two continuous transitions instead of
 the  single first order melting  transition of  three dimensional crystals.
 A hexatic  phase appears between ordered (solid)
and isotropic (liquid) phases.  We shall
refer to this theory as KTHNY theory. The crystalline phase has both long-range
crystalline order and bond orientational order, but the crystalline order
is lost at the 
transition to the hexatic phase when according to KTHNY theory dislocation
pairs unbind. As the temperature is raised further 
a  second continuous transition occurs above which  orientational order is
lost when disclination pairs unbind and the isotropic liquid state forms.
The correlation length  $\xi$ over which there is short-range 
crystalline order diverges
as $T$ approaches the melting temperature. It is given by the expression:
\begin{equation}
\xi(T)\propto \exp\left(\frac{b}{(T/T_m-1)^\nu}\right)
\label{correlation}
\end{equation}
where $b$ is a numerical factor and $\nu=0.36963\ldots$

The KTHNY scenario is not the only mechanism possible for two-dimensional
melting. It may happen that 
a single first order  transition  occurs before the  defects unbind.
 Experimentally, there are systems to which the KTHNY theory applies
 and there are systems where just a single first order transition takes place.
  Electrons on the surface of helium
\cite{GA88},
 submicron polymer colloids confined between glass plates \cite{MW87},
structural order of two dimensional charge--density-waves in
Nb$_x$Ta$_{1-x}$S$_2$ \cite{DL92}  and
 colloidal particles  confined to a monolayer 
 with dipole interactions
\cite{KM94} are examples of the
former while xenon on graphite \cite{JB89} is a example of the latter.

However, Monte Carlo (MC) and molecular dynamics (MD) simulations of two
 dimensional melting  almost always  seem
 to indicate that melting takes place via
 a first order transition \cite{BG82,TC82,ZC92,LS92}. Recently, there has been
a very large scale  simulation \cite{BA96} of
a system of particles interacting via a $r^{-12}$ potential with periodic
boundary conditions.
 Systems of 4096, 16384 and 65536 particles were studied and as the number of 
particles increased the size  of the  van der Waals 
  loops  decreased implying that in the thermodynamic limit the melting
transition might be continuous. It is the chief purpose of this paper
to point out that it is much easier to get results valid in the thermodynamic 
limit by making the two-dimensional system live on the surface of a sphere. 

 Problems arise with MC and MD simulations  when the  relaxation times in
the system become comparable
with the timescale of the simulation.
 When plotting pressure versus density along an isotherm,
a hysteresis loop can appear around the phase transition region. This is
expected for  systems with a first order transition but a system out of 
 equilibrium can  also show hysteresis. It has been found that
deliberately  introducing vacancies into the simulations with periodic 
boundary conditions decreases the jump at the first order transition
 \cite{ST88}. This suggests that one of the problems of doing simulations with 
periodic boundary conditions is that the timescales employed in these
  simulations might be
less than the timescale for nucleating defects such as vacancies,
dislocation pairs etc. thus preventing the attainment of true equilibrium. Now
on the surface of a sphere the crystalline state always contains at least 12
disclinations (i.e 5-fold coordinated sites)  due to Euler's theorem (see eg.
\cite{PD97}).
 Furthermore,  dislocations appear even in the
ground state to screen the 12 disclinations  in order to reduce their
 elastic strain
energy  \cite{PD97}. In other words, the topology of the sphere
always forces a number of defects into the crystalline state. 
But instead of being a disadvantage as one might have first thought, the
presence of the defects seems to allow the system to relax more readily
 and so explore more phase space on the timescale of the simulation. It is
possible that this is the mechanism which makes simulations on
 the surface of a sphere closer to the thermodynamic limit 
  than those done with 
 periodic boundary conditions 

Thermodynamic properties such as the free
energy per particle 
for  particles with short range interactions are the same on the flat plane and 
the sphere as  $N\longrightarrow \infty$. 
When $N\longrightarrow \infty$ the local curvature
vanishes ($R$, the radius of the sphere is varied with $N$ so as to keep
the surface density  $\rho$ constant) so that locally the system on the sphere  
appears flat. 

We have studied by means of MD a system of $N$ particles constrained to
move on the surface of a sphere. Particles interact with each other by
the  repulsive pair potential
\begin{equation}
v(r_{ij})=\epsilon(\sigma/r_{ij})^{12}
\end{equation}
The distance $r_{ij}$ between particles $i,j$ is measured along the geodesic
truncating at $r_{ij}=3\sigma$.  $r_{ij}$ is given by the expression
\begin{equation}
r_{ij}=\vartheta_{ij}R
\end{equation}
where  $\vartheta_{ij}$ is the angle between  particles $i$ and $j$.
\begin{equation}
\vartheta_{ij}=\arccos \left( \sin \theta_i \sin \theta_j \cos \left(
\phi_i-\phi_j \right) +\cos \theta_i \cos  \theta_j \right)
\end{equation}
In this paper we will use reduced units
($\sigma=\epsilon=m=1$).

We employed a velocity Verlet algorithm. It was adapted to
our particular case of polar coordinates $\theta_i,\phi_i$ with
$0\leq\theta_i\leq \pi$ and $0\leq \phi_i< 2\pi$ for $i=1,2,\ldots N$;
\begin{eqnarray}
\theta_i(t+\delta t) &= &\theta_i(t)+(v_{\theta_i}(t)\delta t
+1/2a_{\theta_i}(t)\delta t^2)R^{-1}
\nonumber\\
\label{verlet}\\
\phi_i(t+\delta t)& =& \phi_i(t)+\frac{2(v_{\phi_i}(t)\delta t
+1/2a_{\phi_i}(t)\delta t^2)}
{\sin\theta_i(t+\delta t)+\sin\theta_i(t)}R^{-1}   \nonumber
\end{eqnarray}
where $a_{\phi_i}=\dot{v}_{\phi_i}, a_{\theta_i}=\dot{v}_{\theta_i},
v_{\phi_i}=R\sin\theta_i \dot{\phi_i}, v_{\theta_i}=R\dot{\theta_i}$.
Accelerations $a_{\phi_i}, a_{\theta_i}$ for this problem are as follows
\begin{eqnarray}
a_{\theta_i}& = & -\frac{12}{R^{13}}\sum_{j\neq i}^N \frac{\cos\left(
\phi_i-\phi_j\right)\cos\theta_i\sin\theta_j-\cos\theta_i\cos\theta_j}
{\vartheta_{ij}^{13}\sqrt{1-(\cos\vartheta_{ij}})^2} \nonumber \\
\label{aceleraciones} \\
a_{\phi_i}& = &\frac{12}{R^{13}}\sum_{j\neq i}^N \frac{\sin\left(\phi_i-\phi_j
\right)\sin\theta_j}
{\vartheta_{ij}^{13}\sqrt{1-(\cos\vartheta_{ij}})^2} \nonumber
\end{eqnarray}

Temperature is introduced in the simulation by 
 reselecting the velocities of all the 
particles at once according to a Boltzmann distribution. This reselection
is done at equally spaced intervals of time\cite{AS83}.

 It seems natural to 
choose values for $N$ such   that the particles can be disposed on
the sphere  in a triangular--like lattice. This occurs  when $N=10T+2$
where $T$ is equal to
\begin{equation}
T=h^2+hk+k^2
\label{T}
\end{equation}
with $h$ and $k$ integers \cite{CK62,C93}. For values of $N$'s satisfing
 Eq. (\ref{T}) the particles  can be
arranged on the surface of the sphere into a state with 
full icosahedral symmetry. In this paper we study
 systems of 72, 122 and 272 particles.

It easy to see that the calculation of the accelerations from 
Eq. (\ref{aceleraciones}) is  slower than the corresponding
 calculations
on a flat plane as each  element in the sum in (\ref{aceleraciones})
requires the (slow) computation of several trigonometric functions. This limits
our simulation to rather small values of $N$, so it is fortunate
  that on the sphere results consistent with KTHNY theory can be seen with
 small values of $N$.
The obvious idea of using look--up tables for the trigonometric functions
led to considerable errors.  We expect though that
improvements in our  numerical procedures  can be found. This  would allow
us to study larger systems and so see the hexatic phase, which because it
exists only in a narrow density region  above the melting density,
 is hard to disentangle from  finite size 
effects in small systems.

\begin{figure}
\epsfxsize=\hsize
\begin{center}
\leavevmode
\epsfbox{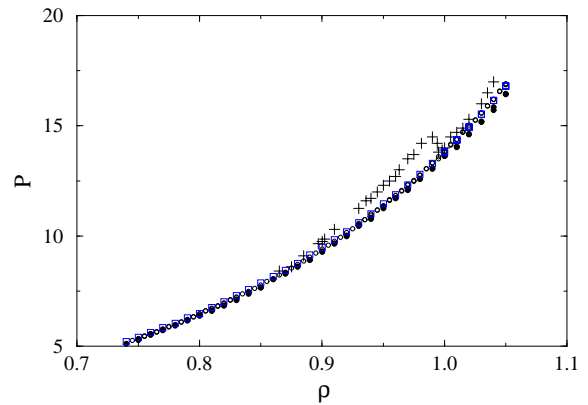}
\end{center}
\caption{Pressure--density plot for 72 (solid circles), 122 (empty circles) and
272
(empty squares) particles on a sphere.
 Results from Broughton {\it et al.\/} (crosses) 
 for the same potential on a flat
 plane are also plotted for comparison.}
\end{figure}

We have calculated the  pressure $P$ and the structure factor 
along the isotherm $T=1$. A simulation time   400,000$\delta t$ was used 
 for each value of the density, where
$\delta t=0.005(m\sigma^2/\epsilon)^{1/2}$.
The pressure was  evaluated using the expression \cite{AT87}
\begin{equation}
P=\rho k_{\rm B}T+\frac{6}{N}\sum_{i<j}^N\langle r_{ij}^{-12} \rangle
\end{equation}
Fig. 1 shows the pressure--density
isotherm for 72 (solid circles),  122 (empty circles) and 272 (empty squares).
Crosses represent data
obtained by Broughton {\it et al.\/}\cite{BG82} for the same interacting
potential on a flat plane. At low densities, the  pressures are the same for the
flat plane
and for the sphere when  both systems are closer to ideal
gas behavior.
As one can see in Fig. 1, there is no hysteresis loop on the sphere
indicating  that the solid phase  does not
melt by a first order transition.

If one concedes that there is no evidence for a first order melting transition
then one might try to argue that what is happenning instead is that the
topology
of the sphere is preventing a solid phase forming and that the system is always 
liquid. However, by studying the structure factor we can see that a transition
to a crystalline phase does indeed occur on the sphere.

The structure factor $S(k)$ is the Fourier transform of the pair correlation
function $h(r)$.
$h(r)$ is related to the pair distribution function $g(r)$ simply by
 $h(r)=g(r)-1$.
Several authors have developed methods to calculate the structure factor in
spherical geometries
\cite{OM93,HL79}. We obtained it from the pair distribution function using the
equation
\begin{equation}
S(k)=1+2\pi\rho R^2 \int_0^\pi h(R\theta )\sin\theta J_0(kR\theta )d\theta
\end{equation}
$J_0$ is the Bessel function of zeroth order.
 Fig. 2 shows the
structure factor for $\rho =.9$. The  dashed line is a Lorentzian fit  to
the first  peak of the structure factor (which occurs at a wave-vector
corresponding to the first reciprocal lattice vector of a triangular lattice).
We define the inverse correlation length as the width of the first peak. Fig. 3
shows the behavior of the inverse width as a function of the density.
 
\begin{figure}
\epsfxsize=\hsize
\begin{center}
\leavevmode
\epsfbox{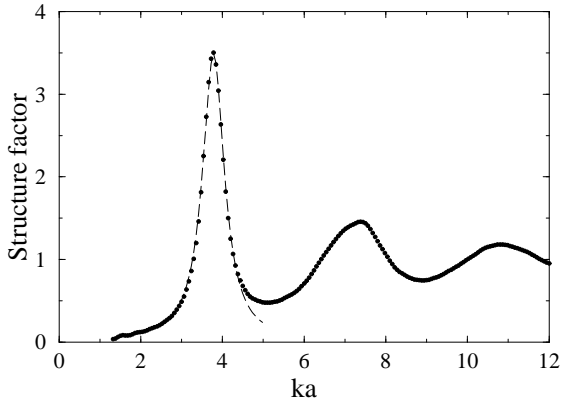}
\end{center}
\caption{Structure factor for $\rho=0.9$ (solid circles). Dashed line is the 
 Lorentzian fit
 used to obtain the  width of the first peak. $a$ is a measure of the 
 particle separation where $a=(\pi\rho)^{-1/2}$}
\end{figure}

\begin{figure}
\epsfxsize=\hsize
\begin{center}
\leavevmode
\epsfbox{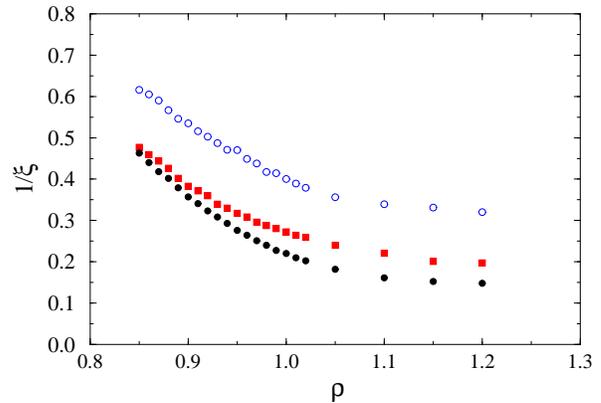}
\end{center}
\caption{Width of the first peak of the structure factor, that is
the inverse correlation length, as a function of density $\rho$.
$N$ is equal to 72 (empty circles), 122 (solid squares) and 272 (solid circles)
}
\end{figure}

$\xi$ saturates
when  $\rho \approx 1$ when $\xi$ is of the order of the radius
$R$ of the sphere.
This kind of behavior is as expected for a continuous  phase
 transition.

Eq. (\ref{correlation}) gives the dependence of $\xi$ in the liquid
phase along an isochore. The dependence of $\xi$ along an isotherm
 is given by
\begin{equation}
\xi(\rho)\propto \exp\left( \frac{b}{\left((\rho_m/\rho)^6-1
\right)^\nu}\right)
\label{iso}
\end{equation}
where $\nu=0.36963$ as before and $\rho_m$ is the melting density.

To obtain Eq. (\ref{iso}) from Eq. (\ref{correlation})
 we have used a scaling properties of the
$r^{-12}$ potential \cite{BG82};  viz that systems with different $T$'s
 and $\rho$'s share the same
thermodynamic properties if they have the same value of $\Gamma$, where 
 $\Gamma=(\pi\rho)^6/(k_{\rm B}T)$.

Fig. 4 shows a log--log plot of $\ln \xi$ as a function of
$(1/\rho^6-1)^{0.36963}$.  (We have taken $\rho_m=1$).
 A straight line of slope equal to -1 is also shown. This
slope is the KTHNY prediction. The plot 
should thus should have  a slope of -1 for densities
approaching $\rho_m$ from below. (Very close to $\rho_m$ finite size effects
will modify the behavior).  The results obtained are clearly consistent
with KTHNY theory, Eq. (\ref{iso}).

\begin{figure}
\epsfxsize=\hsize
\begin{center}
\leavevmode
\epsfbox{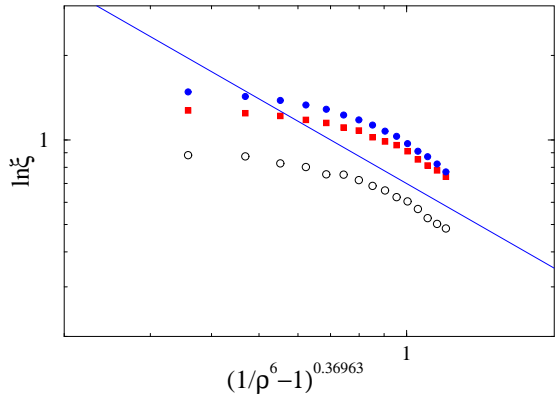}%
\end{center}
\caption{ln$\xi$  versus $(1/\rho^6-1)^{0.36963}$. The slope of the straight
line is -1 which corresponds to KTHNY predictions.
$N$ is equal to 72 (empty circles), 122 (solid squares) and 272 (solid
circles)}
\end{figure}

To summarize, we have simulated a system of classical interacting particles
on a sphere interacting via a short--range interaction.
In this geometry there is no hysteresis loop as on a flat plane. By studying the
 structure factor, evidence of a continuous  melting transition was found 
 near $\rho=1$. The
correlation length of short--range crystalline order in the liquid phase
 diverged on approaching the transition as predicted by
KTHNY theory. We believe that this constitutes strong evidence that simulations
of two dimensional melting phenomena are best performed on the surface of a
sphere and implies that the first order transition so often
reported in simulations on the flat plane is nothing but a numerical artifact. 

We would like to acknowledge financial support from the Direcci\'on
General de Investiga\-ci\'on Cien\-t\'{\i}\-fica y T\'ec\-nica, project number
PB
93/1125, and a grant for APG. We also aknowledge useful discussions with
A. Somoza and M. Ortu\~no.

\end{document}